\input{aipcheck}

\documentclass[
    ,final            
  ,sort&compress
 ,numberedheadings 
  ]
  {aipproc}

\layoutstyle{6x9}

\newcommand\aj{AJ} 
\newcommand\apj{{ApJ}} 
\newcommand\apjl{{ApJL}} 
\newcommand\aap{{A\&A}} 
\newcommand\pasp{{PASP}} 

\newcommand{\kms}{km~s$^{-1}$}

\newcommand{\lbol}{$\log_{10}{L_{bol}/L_{\odot}}$}
\newcommand{\teff}{T$_{eff}$}
\newcommand{\logg}{$\log{g}$}

\begin{document}

\title{Ultracool Subdwarfs: The Halo Population Down to the Substellar Limit}


\author{Adam J. Burgasser}{
  address={Massachusetts Institute of Technology, Cambridge, MA, USA}
}

\author{Sebasti\'en L\'epine}{
  address={American Museum of Natural History, New York, NY, USA}
}

\author{Nicolas Lodieu}{
  address={Instituto de Astrofisica de Canarias, La Laguna, Tenerife, Spain}
}

\author{Ralf-Dieter Scholz}{
  address={Astrophysical Institute Potsdam, Potsdam, Germany}
}

\author{Phillippe Delorme}{ 
  address= {Laboratoire d'Astrophysique de Grenoble,
                Observatoire de Grenoble, Grenoble,
                France}
}

\author{Wei-Chun Jao}{
  address={Georgia State University, Atlanta, GA, USA} }

\author{Brandon J.\ Swift}{
  address={Steward Observatory, University of Arizona, Tucson, AZ, USA} }

\author{Michael C.\ Cushing}{
  address= {Institute for Astronomy, University of Hawai'i, Honolulu, HI, USA}
}

\begin{abstract}
Ultracool subdwarfs are low luminosity, late-type M and L dwarfs that exhibit spectroscopic indications of subsolar metallicity and halo kinematics.  Their recent discovery and ongoing investigation have led to new insights into the role of metallicity in the opacity structure, chemistry (e.g. dust formation) and evolution of low-temperature atmospheres; the long-term evolution of magnetic activity and angular momentum amongst the lowest-mass stars; the form of the halo luminosity and mass functions down to the hydrogen-burning mass limit; and even fundamental issues such as spectral classification and absolute brightness scales. This Splinter Session was devoted to bringing advances in observational and theoretical ultracool subdwarf research  to the attention of the low-mass stellar and brown dwarf communities, as well as to share results among ultracool subdwarf enthusiasts.
\end{abstract}

\maketitle


\section{Introduction}

The late-type M and L dwarf classes of ultracool dwarf stars and brown dwarfs are now well-sampled, due in large part to wide-field red and infrared imaging surveys such as 2MASS, DENIS and SDSS.  We are now beginning to uncover the metal-poor, halo counterparts to these disk dwarfs, the so-called {\bf ultracool subdwarfs} \cite{2005ESASP.560..237B}.  These sources have been found primarily in wide-field imaging surveys and the first generations of red/near-infrared proper motion surveys.  Discoveries include the first L-type---and possibly T-type---subdwarfs, extending our knowledge of the Galactic halo down to the substellar (brown dwarf) regime.  As the low-mass ultracool subdwarfs, with their extremely long nuclear burning lifetimes, were presumably formed early in the Galaxy's history, they are important tracers of Galactic structure and chemical enrichment history.  In addition, detailed studies of their complex spectral energy distributions are facilitating new insights on the role of metallicity in the opacity structure, chemistry and evolution of cool atmospheres; the long-term evolution of magnetic activity and angular momentum; and fundamental issues of spectral classification and temperature/luminosity scales.

This Splinter was devoted to highlighting advances in ultracool subdwarf research, both observational and theoretical.  It was organized into three subtopics: Discoveries, Classification and  Fundamental Parameters.  Individual presentations will be made available at \url{http://www.browndwarfs.org/cs15}.

\section{Discoveries}

\subsection{An Extremely Wide and Very Low-Mass Common Proper
   Motion Pair - Representatives of a Nearby Halo Stream? \\ (R.-D.\ Scholz et al.)}
   
\begin{figure}
  \includegraphics[height=.48\textheight, angle=-90]{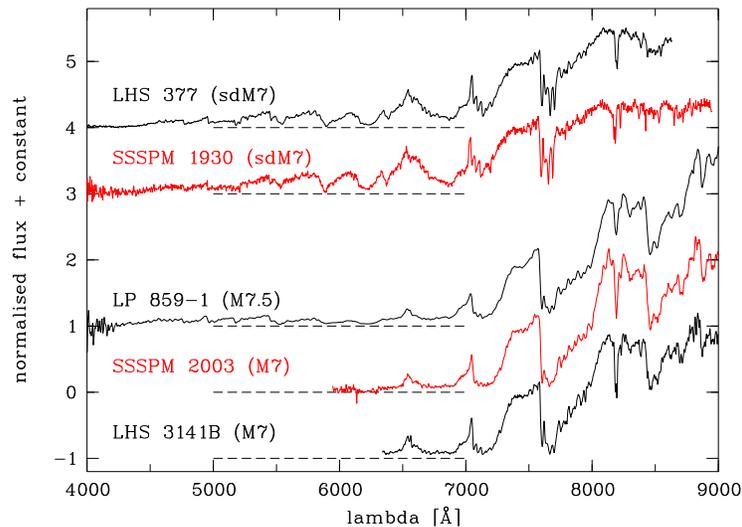}
  \caption{Low-resolution classification spectra of SSSPM\,J2003$-$4433 (M7)
and SSSPM\,J1930$-$4311 (sdM7) and some comparison sources
(from \cite{2008A&A...487..595S}).
\label{fig_scholz}}
\end{figure}

A pair of faint stars sharing exactly the same very large
proper motion of about 860 mas/yr and separated by about
six degrees has been discovered in a high proper motion
survey of the southern sky using multi-epoch positions and
photometry from the SuperCOSMOS Sky Surveys. The two stars,
SSSPM\,J2003$-$4433 and SSSPM\,J1930$-$4311, have been
classified as a late-type (M7) dwarf and an ultracool subdwarf
(sdM7) (Figure~\ref{fig_scholz}) with individually estimated spectroscopic
distances of 38 pc and 72 pc, respectively. In view of
the accurate agreement in their large proper motions a common
distance of about 50 pc and a projected physical separation
of about 5 pc has been assumed, ruling out a physical binary.
The mean heliocentric space velocity of the pair
($U,V,W$) = ($-232,-170,+74$) {\kms} is typical of the Galactic
halo population. These values rely on a preliminary radial velocity
measurement and on the assumption of a common distance and
velocity vector. The large separation and the different
metallicities of dwarf and subdwarf make a common formation
scenario as a wide binary (later disrupted) improbable.
It seems more likely that this wide pair is part of an old
halo stream \cite{2008A&A...487..595S}.

\subsection{A Metal-poor Mid-T dwarf from the CFBDS Survey \\ (P.\ Delorme et al.)}

We report the discovery of CFBDS~J150000-182407, 
a T subdwarf candidate which we identified
during the Canada France Brown Dwarf Survey (CFBDS).
CFDBS \cite{2008A&A...484..469D} is an $i'-z'$ wide-field search
for ultracool brown dwarfs which uses the MegaCam wide-field camera  on the
Canada-France-Hawaii Telescope (CFHT).

\begin{figure}
\includegraphics[width=5.3cm,angle=-90]{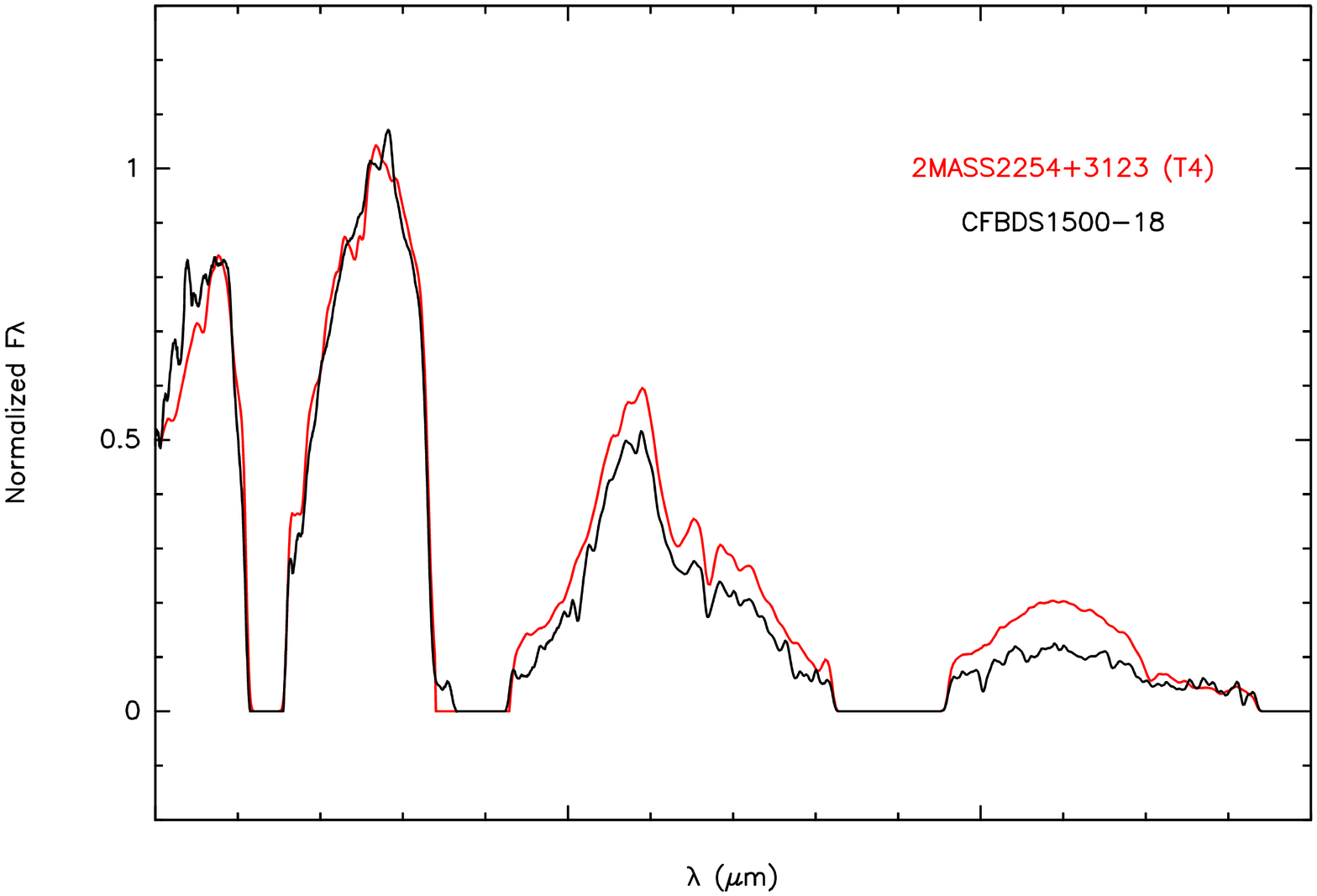}
\includegraphics[width=5.3cm,angle=-90]{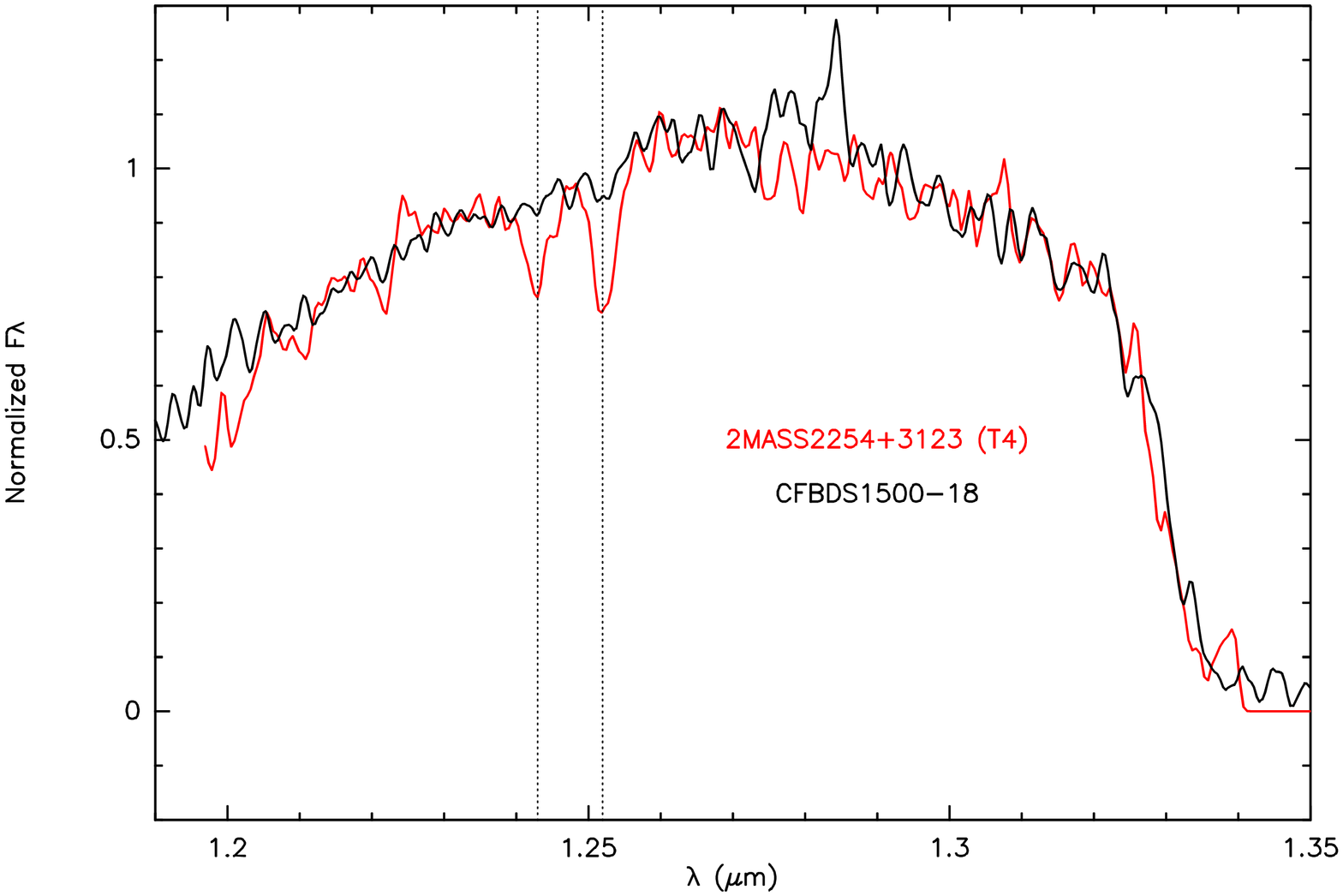}
\caption{Spectra of T subdwarf candidate CFBDS1500 (black line) and 2M2254 (red line, from  \cite{2008ApJ...678.1372C}), the latter a solar-metallicity T dwarf template that best matches the spectrum of
  CFBDS1500. Left panel shows the whole NIR spectrum while the right panel zooms in on $J$-band region. The vertical lines mark the locations of 
the K~I doublet.}
\label{spectre_comp2}
\end{figure}

CFBDS1500 is a peculiar T4.5 dwarf, with strong spectroscopic evidence 
for a subsolar metallicity. Comparison of the overall shape of 
its spectrum with synthetic spectra from \cite{2006ApJ...640.1063B} suggests [M/H]$\sim$-0.3,
while the complete absence of the $\sim$1.25$\mu$m K~I doublet (Figure~\ref{spectre_comp2}) 
rather suggests that [M/H]$\leq$-0.5 when comparing with the same models. The kinematics of CFBDS1500 
imply an 80\% probability that it belongs to the thick disk, but 
leaves $\sim$10\% probability that it instead is an older member
of the thin disk or a member of the halo. [M/H]$\sim$-0.5 is consistent with
either thick disc membership or an older thin disk population,
while [M/H]$\sim$-0.3 would lean slightly towards the older
thin disk but remains easily compatible with thick disk membership.

Spectroscopy and kinematics together make CFBDS1500 a strong
thick disk candidate, and as such the first incontrovertibly 
substellar object which would not belong to the thin disk. 
As one of the few low-metallicity brown dwarfs known, 
it is a major benchmark for synthetic spectra. The inconsistency between the
metallicity determined from the strength of the K~I doublet
and from the general shape of the spectrum will need to be 
explained. We tend to trust the metallicity constrain from 
the K~I doublet more, since it depends (besides, admittedly, 
the structure of the atmosphere) on a single line pair
with very well understood atomic physics, rather than on millions
of incompletely characterized molecular lines. 

 If the [M/H]$\sim$-0.5 metallicity holds, the
observed spectrum would suggest that the models overestimate the
sensitivity of the general shape of the spectrum to metallicity, 
especially in the $J$-band. That would hardly be a surprise,
given the many poorly constrained physical inputs that need
to be taken into account. 
Since the discovery of brown dwarfs belonging to old Galactic
populations is now possible, it  becomes crucial to interpret their
spectra with state-of-the-art evolutionary models. The only one for low metallicity 
brown dwarfs (to our knoweldge) was calculated
without taking into account clouds in the atmosphere \cite{2001RvMP...73..719B}. New theoretical
work on this topic (e.g., \cite{2008arXiv0808.2611S}) is now a critical point of T sub-dwarfs characterization.

\section{Classification}

\subsection{A Spectral Sequence of K- and M-type Subdwarfs \\ (W.-C.\ Jao et al.)}

Using new spectra of 88 K and M-type subdwarfs, we have considered novel
methods for assigning spectral types and take steps toward
developing a comprehensive spectral sequence for subdwarf types K3.0
to M6.0.  The types are assigned based on the overall morphology of
spectra covering 6000\AA~to 9000\AA~through the understanding of GAIA
model grids. The types and sequence presented link the spectral types
of cool subdwarfs to their main sequence counterparts, with emphasis
on the relatively opacity-free region from 8200--9000\AA.  When
available, supporting abundance, kinematic, and trigonometric parallax
information is used to provide more complete portraits of the observed
subdwarfs.  We find that the CaHn (n= 1--3) and TiO5 indices often
used for subdwarf spectral typing are affected in complicated ways by
combinations of subdwarfs' temperatures, metallicities, and gravities,
and we use model grids to evaluate the trends in all three parameters.
Because of the complex interplay of these three characteristics, it is
not possible to identify a star as an ``extreme'' subdwarf simply
based on very low metallicity, and we suggest that the modifiers
``extreme'' or ``ultra'' only outline locations on spectroscopic
indices plots, and should not be used to imply low or very low
metallicity stars.  In addition, we propose that ``VI'' be used to
identify a star as a subdwarf, rather than the confusing ``sd''
prefix, which is also used for hot O and B subdwarfs that are
unrelated to the cool subdwarfs.
These results have been published in \cite{2008AJ....136..840J}.

\subsection{Spectroscopic Sequences of Cool and Ultra-cool Subdwarfs (sdM/esdM/usdM) from the Sloan Digital Sky Survey \\ (S.\ L\'epine et al.)}

We present new spectra of cool and ultra-cool subdwarfs,
obtained from the Sloan Digital Sky Survey (SDSS). Thousands of these sources have now been observed in SDSS (see \cite{2008ApJ...681L..33L} and 
poster by
L\'epine, these proceedings). The stars were categorized in a new,
recently expanded classification system \cite{2007ApJ...669.1235L}
as subdwarfs (sdM), extreme subdwarfs (esdM), and
ultrasubdwarfs (usdM), based on the ratio of the TiO to CaH molecular
bands. It is argued that the M/sdM/esdM/usdM subclasses represent
a metallicity sequence (cf.\ \cite{1997PASP..109.1233G,2006PASP..118..218W}), 
with the usdM being the most metal-poor of the
subdwarfs. 

While it has been suggested by others that the morphology of the
subdwarfs depends on both their metallicity and gravity, observational
evidence supports the idea that metallicity is the dominant parameter,
and that the current classification system effectively ranks the stars
according to mass (subtype) and metallicity (subclass). As
demonstrated in Figures 2 and 3 of \cite{2007ApJ...669.1235L}, 
stars kinematically selected from the Galactic disk display only
small variations in their TiO/CaH ratio, while stars kinematically
selected from the Galactic halo display a very broad range of TiO/CaH
values. The disk population has a broader range of stellar ages, and
should also have a broader range of surface gravities than the halo
stars, which are uniformly old. The fact that the disk stars have a
much smaller range of TiO/CaH than the halo stars indicates than
metallicity effects must dominate in the subdwarfs. Furthermore,
atmospheric models indicate that gravity variations not only change
the TiO/CaH ratio, but they also notably affect the strength of atomic
spectral lines such as KI and NaI. However SDSS spectra show no clear
variation in the equivalent widths of the KI, NaI, and CaII lines for
stars of a given subclass and subtype, which again suggest that stars
of similar subtypes have similar surface gravities. Gravity, in the
nearby halo subdwarfs, is not a free parameter but is essentially set
by the initial mass and metallicity of the subdwarf.

Ultra-cool subdwarfs are particularly useful for refining the current
classification scheme, because their molecular bands and atomic lines
are more prominent. Spectra of many cool and ultra-cool subdwarfs are
now available in the SDSS database, and new ones should become
available in the future as halo subdwarfs are now specifically
targeted for follow-up spectroscopy. The new SDSS spectra also have
a broad spectral coverage (4000\AA-9000\AA), and spectra from many
late-type subdwarfs can be combined to obtain high signal-to-noise
spectral templates. A fit to these templates yield a more reliable
classification than the narrowly defined TiO5, CaH2, and CaH3 spectral
indices. The use of the spectral indices in the classification is
useful, but should be phased out in favor of the new SDSS
classification templates (L\'epine, in preparation).

\subsection{L Subdwarfs: Classification, Distance Scale and Low-Metallicity Condensate Formation \\ (A.\ Burgasser et al.)}

\begin{figure}
\includegraphics[width=8cm,angle=0]{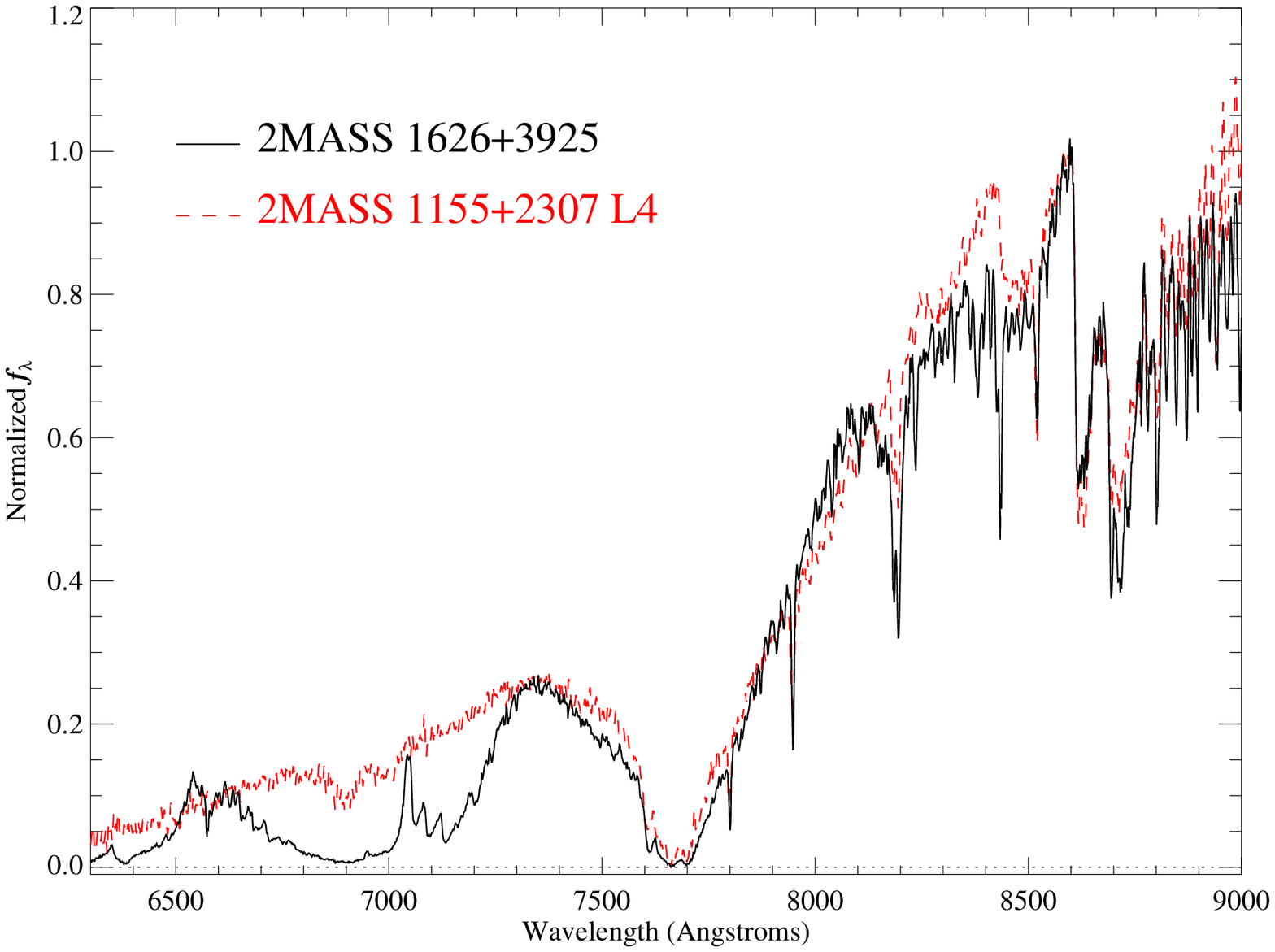}
\includegraphics[width=6cm,angle=0]{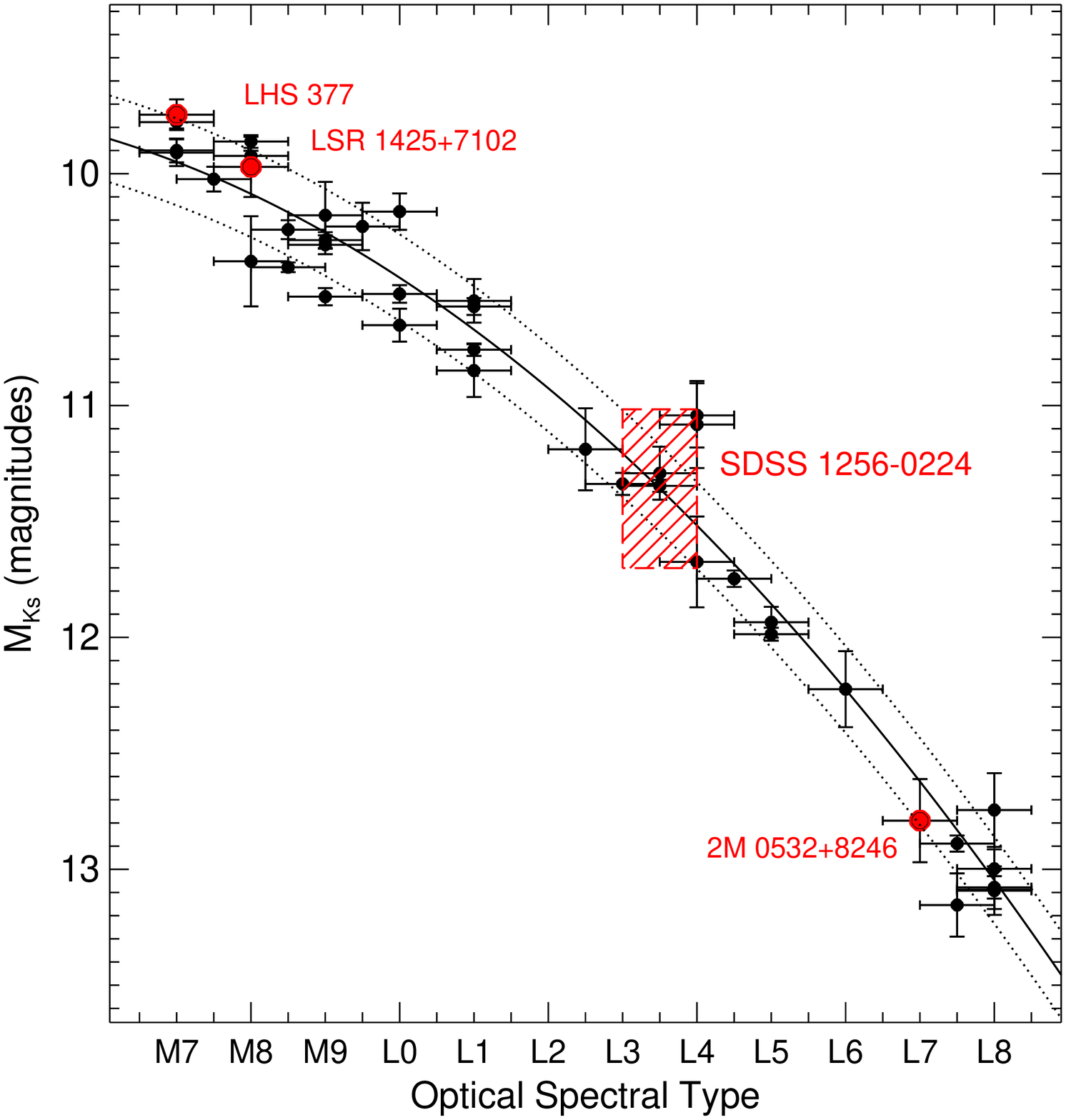}
\caption{{\em Left}: Comparison of the red optical spectra of the sdL4 2MASS~1626+39 to the L4 spectral standard 2MASS~1155+23, illustrating their similarities in the 7300--9000~{\AA} range (from \cite{2007ApJ...657..494B}). {\em Right}: $M_{K_s}$/spectral type relation for late-type M and L dwarfs (black points, polynomial fit indicated by the solid line with dashed lines indicating $\pm$1$\sigma$)
compared to measurements for the sdM7 LHS 377, sdM8 LSR 1425+71 and sdL7 2MASS~0532+82 (red points; \cite{1992AJ....103..638M,2008arXiv0806.2336D, 2008ApJ...672.1159B}).  The predicted $M_{K_s}$ for the sdL3.5 SDSS~1256-02 is indicated by the hatched region (from \cite{bur1256}).
\label{fig_sdl}}
\end{figure}

L subdwarfs are the metal-poor counterparts to the L dwarf class of low mass stars and brown dwarfs, effectively extending our sampling of halo stars to the (metallicity-dependent) hydrogen-burning mass limit.
With strong metal hydride bands, weak metal oxides and red optical spectral energy distributions, the L subdwarfs share many of the same spectral characteristics as L dwarfs, but differ in their considerably bluer near-infrared colors ($J-K_s$ $\approx$ 0 vs 1.5--2.5) and much stronger metal hydride, metal oxide and alkali line absorption in the red optical (Figure~\ref{fig_sdl}). The first L subdwarfs were identified in 2003; today there are at least three reported in the literature \cite{2003ApJ...592.1186B,2004ApJ...614L..73B,siv04}, not including the unusual source LSR 1610-0040 \cite{2003ApJ...591L..49L} discussed in detail below.

One of the recent advances in L subdwarf research is a formalized spectral classification scheme for these objects.  While the classification of M subdwarfs is actively debated due to the large samples now available (see contributions from S.\ L\'{e}pine and W.-C.\ Jao), there are too few L subdwarfs to accurately define the class.  \cite{2007ApJ...657..494B} have proposed that L subdwarfs be classified according to the closest match to the L dwarf standards of \cite{1999ApJ...519..802K} in the 7300--9000~{\AA} range (see Figure~\ref{fig_sdl}), a region in which peculiarities are minimized.    This had provided preliminary types for 
2MASS 0532+82 (sdL7 \cite{2003ApJ...592.1186B}),
2MASS 1626+39 (sdL4 \cite{2004ApJ...614L..73B}) 
and SDSS 1256-02 (sdL3.5 \cite{siv04}).

As the first L subdwarf identified, 2MASS 0532+82 has been the best studied.  It is the first L subdwarf to have its astrometric parallax measured ($\pi$ = 37.5$\pm$1.7~mas \cite{2008ApJ...672.1159B}), and combined with photometry from 2MASS and {\em Spitzer} \cite{2006ApJ...651..502P} we have measured a luminosity of {\lbol} = $-4.24{\pm}0.06$ and estimate {\teff} = 1730$\pm$90~K for this source.  These parameters are comparable to those of mid-type L field dwarfs, suggesting that L subdwarf classifications are ``later'' than corresponding L dwarf types.  Interesting, these values make the hydrogen burning status of 2MASS 0532+82 somewhat ambiguous, since depending on its metallicity (which is largely unconstrained) this source may be a star, a brown dwarf, or a brown dwarf which will ultimately reach the main sequence and become a star!   Examining absolute magnitude relations, it appears that the $M_K$/spectral type relation of L subdwarfs is roughly coincident with those of L dwarfs (Figure~\ref{fig_sdl}), although more astrometric work is needed (see also \cite{2008arXiv0806.2336D}). 2MASS~0532+82 is also the only L subdwarf with high resolution spectral observations \cite{2006AJ....131.1806R}, which indicate rapid rotation ($v\sin{i} = 65{\pm}15$~{\kms}) suggesting the absence of angular momentum evolution for very low mass stars.   The $UVW$ velocities of this source unambiguously confirm it as a halo object, with a retrograde Galactic orbit relative to the Galactic disk. 

We are currently examining theoretical spectral models for the sdL3.5 SDSS~1256-02 (\cite{bur1256}; also see next contribution by B.~Swift).  Using the {\em Phoenix-Drift} models \cite{2008ApJ...675L.105H}, we have attempted to reproduce the observed optical and near-infrared colors and spectra of this source.  Our results so far indicate a {\teff} $\approx$ 2100--2500~K and [M/H] $\approx$ -1.5 -- -1.0~dex, although there remain strong discrepancies particularly in the red optical.  Interestingly, we find that even with the most advanced treatment of (metallicity-dependent) cloud formation, clouds in the models are too thick and condensation too efficient to match the observations of SDSS~1256-02, consistent with the ``cloud suppression'' in ultracool subdwarf atmospheres suggested in prior studies \cite{2003ApJ...592.1186B,2006AJ....131.1806R, 2006AJ....132.2372G,2007ApJ...657..494B}

\section{Fundamental Parameters}

\subsection{Theoretical Modeling of L Subdwarf Spectra \\ (B.\ Swift et al.)}

We present fits of atmospheric models of varying metallicity to the published optical and near-infrared spectrum and IRAC photometry of the sdL7 2MASS J05325346+8246465 \cite{2003ApJ...592.1186B}. This source is the best-observed member of a growing population of what are suspected to be metal-poor ultracool dwarfs, and the only one with a parallax distance. The model selection was made using a goodness-of-fit statistic in conjunction with Monte Carlo simulated data to account for uncertainties in the absolute flux; this statistic was then examined in various bands to study systematic issues in the models and to produce a concordance fit for the properties of this object in {\teff}, {\logg} and [M/H] space.

\subsection{Astrometric Observations of LSR 1610-0040 \\ (M.\ Cushing et al.)}

Since its discovery \cite{2003ApJ...591L..49L}, LSR 1610-0040 has defied
explanation.  Its red optical spectrum and high proper motion suggest
that LSR 1610-0040 is an early-type L subdwarf, yet its near-infrared spectrum
indicates a mid-type dM or sdM, albeit with numerous peculiar
spectral features \cite{2006AJ....131.1797C,2006AJ....131.1806R}.
Based on a comparison of its near-infrared spectrum to that of field M
dwarfs, \cite{2006AJ....131.1797C} assign LSR 1610-0040 a spectral type of
sd/dM6.

We present new astrometric observations of LSR
1610-0040 \cite{2008arXiv0806.2336D}.  At a measured distance of $\sim$32 pc, its position in color/
absolute magnitude diagrams ($M_V$ vs $V-I$, $M_{K_s}$ vs $I-K_s$, $M_{K_s}$
vs $J-K_s$) is consistent with mid-type M dwarfs.  However, its $B-V$ color
is 1.2~mags redder than both M subdwarfs and dwarfs of similar spectral
types.  We speculate that the $B$-band magnitude of LSR 1610-0040 is suppressed
due to enhanced AlH absorption since Al appears over-abundant relative
to solar based on the strength of the 1.313 $\mu$m Al I doublet.

Perhaps most interesting is that the astrometric observations indicate
that LSR 1610-0040 is an unresolved binary with a period of 1.66~yr.  The
photocentric orbit has a moderate eccentricity of 0.44$\pm$0.02, a
semi-major axis of 0.28$\pm$0.01~AU, and an inclination of 83$\pm$1$^o$.
Under the assumption that the secondary contributes little to no light
to the system, and using a near-infrared mass luminosity relation, we
estimate the masses of the components to be M$_A$ = 0.095~M$_{\odot}$ and
M$_B$ = 0.059 to 0.082 M$_{\odot}$.  We speculate that LSR 1610-0040A was originally a
0.05 Msun star with [Fe/H] $\sim$ $-2$ which later accreted $<$0.05~M$_{\odot}$ of
material from a massive AGB star that has undergone hot bottom burning.
Pollution by such material, enhanced in Al and Na and depleted in O,
would then explain the peculiar spectrum of LSR 1610-0040A.  LSR 1610-0040B has
too little mass to be the remnant white dwarf of such a hypothetical AGB
star so the AGB star in question must have been a more distant companion
that has since been lost from the system.


\begin{theacknowledgments}
We thank the conference organizers for providing a forum to discuss this topic.
\end{theacknowledgments}



\bibliographystyle{aipprocl} 

\begin{thebibliography}{10}
\providecommand{\enquote}[1]{``#1''}
\expandafter\ifx\csname url\endcsname\relax
  \def\url#1{\texttt{#1}}\fi
\expandafter\ifx\csname urlprefix\endcsname\relax\def\urlprefix{URL }\fi

\bibitem{2005ESASP.560..237B}
A.~J. {Burgasser}, J.~D. {Kirkpatrick}, and S.~{L{\'e}pine},
  \enquote{{Ultracool subdwarfs: metal-poor stars and brown dwarfs extending
  into the late-type M, L and T dwarf regimes},} in \emph{13th Cambridge
  Workshop on Cool Stars, Stellar Systems and the Sun}, edited by F.~{Favata},
  G.~A.~J. {Hussain}, and B.~{Battrick}, 2005, vol. 560 of \emph{ESA Special
  Publication}, pp. 237.

\bibitem{2008A&A...487..595S}
R.-D. {Scholz}, N.~V. {Kharchenko}, N.~{Lodieu}, and M.~J. {McCaughrean},
  \emph{\aap} \textbf{487}, 595--599 (2008).

\bibitem{2008A&A...484..469D}
P.~{Delorme}, C.~J. {Willott}, T.~{Forveille}, X.~{Delfosse}, C.~{Reyl{\'e}},
  E.~{Bertin}, L.~{Albert}, E.~{Artigau}, A.~C. {Robin}, F.~{Allard},
  R.~{Doyon}, and G.~J. {Hill}, \emph{\aap} \textbf{484}, 469--478 (2008).

\bibitem{2006ApJ...640.1063B}
A.~{Burrows}, D.~{Sudarsky}, and I.~{Hubeny}, \emph{\apj} \textbf{640},
  1063--1077 (2006).

\bibitem{2008ApJ...678.1372C}
M.~C. {Cushing}, M.~S. {Marley}, D.~{Saumon}, B.~C. {Kelly}, W.~D. {Vacca},
  J.~T. {Rayner}, R.~S. {Freedman}, K.~{Lodders}, and T.~L. {Roellig},
  \emph{\apj} \textbf{678}, 1372--1395 (2008).

\bibitem{2001RvMP...73..719B}
A.~{Burrows}, W.~B. {Hubbard}, J.~I. {Lunine}, and J.~{Liebert}, \emph{Reviews
  of Modern Physics} \textbf{73}, 719--765 (2001).

\bibitem{2008arXiv0808.2611S}
D.~{Saumon}, and M.~S. {Marley}, \emph{ArXiv e-prints} \textbf{808} (2008).

\bibitem{2008AJ....136..840J}
W.-C. {Jao}, T.~J. {Henry}, T.~D. {Beaulieu}, and J.~P. {Subasavage},
  \emph{\aj} \textbf{136}, 840--880 (2008).

\bibitem{2008ApJ...681L..33L}
S.~{L{\'e}pine}, and R.-D. {Scholz}, \emph{\apjl} \textbf{681}, L33--L36
  (2008).

\bibitem{2007ApJ...669.1235L}
S.~{L{\'e}pine}, R.~M. {Rich}, and M.~M. {Shara}, \emph{\apj} \textbf{669},
  1235--1247 (2007).

\bibitem{1997PASP..109.1233G}
J.~{Gizis}, and I.~{Reid}, \emph{\pasp} \textbf{109}, 1232-1236 (1997).

\bibitem{2006PASP..118..218W}
V.~M. {Woolf}, and G.~{Wallerstein}, \emph{\pasp} \textbf{118}, 218--226
  (2006).

\bibitem{2003ApJ...592.1186B}
A.~J. {Burgasser}, J.~D. {Kirkpatrick}, A.~{Burrows}, J.~{Liebert}, I.~N.
  {Reid}, J.~E. {Gizis}, M.~R. {McGovern}, L.~{Prato}, and I.~S. {McLean},
  \emph{\apj} \textbf{592}, 1186--1192 (2003).

\bibitem{2004ApJ...614L..73B}
A.~J. {Burgasser}, \emph{\apjl} \textbf{614}, L73--L76 (2004).

\bibitem{siv04}
T.~{Sivarani}, A.~K. {Kembhavi}, and J.~{Gupchup}, \emph{ArXiv e-prints}
  (2004).

\bibitem{2003ApJ...591L..49L}
S.~{L{\'e}pine}, R.~M. {Rich}, and M.~M. {Shara}, \emph{\apjl} \textbf{591},
  L49--L52 (2003).

\bibitem{2007ApJ...657..494B}
A.~J. {Burgasser}, K.~L. {Cruz}, and J.~D. {Kirkpatrick}, \emph{\apj}
  \textbf{657}, 494--510 (2007).

\bibitem{1999ApJ...519..802K}
J.~D. {Kirkpatrick}, I.~N. {Reid}, J.~{Liebert}, R.~M. {Cutri}, B.~{Nelson},
  C.~A. {Beichman}, C.~C. {Dahn}, D.~G. {Monet}, J.~E. {Gizis}, and M.~F.
  {Skrutskie}, \emph{\apj} \textbf{519}, 802--833 (1999).

\bibitem{1992AJ....103..638M}
D.~G. {Monet}, C.~C. {Dahn}, F.~J. {Vrba}, H.~C. {Harris}, J.~R. {Pier}, C.~B.
  {Luginbuhl}, and H.~D. {Ables}, \emph{\aj} \textbf{103}, 638--665 (1992).

\bibitem{2008arXiv0806.2336D}
C.~C. {Dahn}, H.~C. {Harris}, S.~E. {Levine}, T.~{Tilleman}, A.~K.~B. {Monet},
  R.~C. {Stone}, H.~H. {Guetter}, B.~{Canzian}, J.~R. {Pier}, W.~I. {Hartkopf},
  J.~{Liebert}, and M.~{Cushing}, \emph{ArXiv e-prints} \textbf{806} (2008).

\bibitem{2008ApJ...672.1159B}
A.~J. {Burgasser}, F.~J. {Vrba}, S.~{L{\'e}pine}, J.~A. {Munn}, C.~B.
  {Luginbuhl}, A.~A. {Henden}, H.~H. {Guetter}, and B.~C. {Canzian},
  \emph{\apj} \textbf{672}, 1159--1166 (2008).

\bibitem{bur1256}
A.~J. {Burgasser}, S.~{Witte}, C.~{Helling}, and P.~H. {Hauschildt},
  \emph{\apj}  (in preparation).

\bibitem{2006ApJ...651..502P}
B.~M. {Patten}, J.~R. {Stauffer}, A.~{Burrows}, M.~{Marengo}, J.~L. {Hora},
  K.~L. {Luhman}, S.~M. {Sonnett}, T.~J. {Henry}, D.~{Raghavan}, S.~T.
  {Megeath}, J.~{Liebert}, and G.~G. {Fazio}, \emph{\apj} \textbf{651},
  502--516 (2006).

\bibitem{2006AJ....131.1806R}
A.~{Reiners}, and G.~{Basri}, \emph{\aj} \textbf{131}, 1806--1815 (2006).

\bibitem{2008ApJ...675L.105H}
C.~{Helling}, M.~{Dehn}, P.~{Woitke}, and P.~H. {Hauschildt}, \emph{\apjl}
  \textbf{675}, L105--L108 (2008).

\bibitem{2006AJ....132.2372G}
J.~E. {Gizis}, and J.~{Harvin}, \emph{\aj} \textbf{132}, 2372--2375 (2006).

\bibitem{2006AJ....131.1797C}
M.~C. {Cushing}, and W.~D. {Vacca}, \emph{\aj} \textbf{131}, 1797--1805 (2006).

\end{thebibliography}


\end{document}